\documentclass[aps,prd,twocolumn,showpacs,nofootinbib]{revtex4-1}

\usepackage{amssymb} \usepackage{amsmath} \usepackage{graphicx}
\usepackage{epsfig,latexsym}
\usepackage{mathtools}% http://ctan.org/pkg/mathtools
\RequirePackage{xspace} \allowdisplaybreaks

\begin{document}

\def\bef{\begin{figure}}
\def\eef{\end{figure}}

\newcommand{\nl}{\nonumber\\}

\newcommand{\ans}{ansatz }
\newcommand{\be}[1]{\begin{equation}\label{#1}}
\newcommand{\beq}{\begin{equation}}
\newcommand{\ee}{\end{equation}}
\newcommand{\beqn}[1]{\begin{eqnarray}\label{#1}}
\newcommand{\eeqn}{\end{eqnarray}}
\newcommand{\bd}{\begin{displaymath}}
\newcommand{\ed}{\end{displaymath}}
\newcommand{\mat}[4]{\left(\begin{array}{cc}{#1}&{#2}\\{#3}&{#4}
\end{array}\right)}
\newcommand{\matr}[9]{\left(\begin{array}{ccc}{#1}&{#2}&{#3}\\
{#4}&{#5}&{#6}\\{#7}&{#8}&{#9}\end{array}\right)}
\newcommand{\matrr}[6]{\left(\begin{array}{cc}{#1}&{#2}\\
{#3}&{#4}\\{#5}&{#6}\end{array}\right)}
\newcommand{\cvb}[3]{#1^{#2}_{#3}}
\def\lsim{\raise0.3ex\hbox{$\;<$\kern-0.75em\raise-1.1ex
e\hbox{$\sim\;$}}}
\def\gsim{\raise0.3ex\hbox{$\;>$\kern-0.75em\raise-1.1ex
\hbox{$\sim\;$}}}
\def\abs#1{\left| #1\right|}
\def\simlt{\mathrel{\lower2.5pt\vbox{\lineskip=0pt\baselineskip=0pt
           \hbox{$<$}\hbox{$\sim$}}}}
\def\simgt{\mathrel{\lower2.5pt\vbox{\lineskip=0pt\baselineskip=0pt
           \hbox{$>$}\hbox{$\sim$}}}}
\def\unity{{\hbox{1\kern-.8mm l}}}
\newcommand{\eps}{\varepsilon}
\def\ep{\epsilon}
\def\ga{\gamma}
\def\Ga{\Gamma}
\def\om{\omega}
\def\omp{{\omega^\prime}}
\def\Om{\Omega}
\def\la{\lambda}
\def\La{\Lambda}
\def\al{\alpha}
\newcommand{\ov}{\overline}
\renewcommand{\to}{\rightarrow}
\renewcommand{\vec}[1]{\mathbf{#1}}
\newcommand{\vect}[1]{\mbox{\boldmath$#1$}}
\def\tm{{\widetilde{m}}}
\def\mcirc{{\stackrel{o}{m}}}
\newcommand{\Dm}{\Delta m}
\newcommand{\dm}{\varepsilon}
\newcommand{\tanb}{\tan\beta}
\newcommand{\nbar}{\tilde{n}}
\newcommand\PM[1]{\begin{pmatrix}#1\end{pmatrix}}
\newcommand{\up}{\uparrow}
\newcommand{\down}{\downarrow}
\def\omE{\omega_{\rm Ter}}

%
%%%%%%%%%%     mauri    %%%%%%%%%%%%%%%%%%%%%%%%%%%%%%%%%

\newcommand{\Dsusy}{{susy \hspace{-9.4pt} \slash}\;}
\newcommand{\DCP}{{CP \hspace{-7.4pt} \slash}\;}
\newcommand{\mc}{\mathcal}
\newcommand{\gr}{\mathbf}
\renewcommand{\to}{\rightarrow}
\newcommand{\gtc}{\mathfrak}
\newcommand{\wh}{\widehat}
\newcommand{\br}{\langle}
\newcommand{\kt}{\rangle}

%%%%%%%%%%%%%%%%%%%%%%%%%%%%%%%%%%%%%%%%%%%%%%%%%%%%%%%%%%

\def\lsim{\mathrel{\mathop  {\hbox{\lower0.5ex\hbox{$\sim$}
\kern-0.8em\lower-0.7ex\hbox{$<$}}}}}
\def\gsim{\mathrel{\mathop  {\hbox{\lower0.5ex\hbox{$\sim$}
\kern-0.8em\lower-0.7ex\hbox{$>$}}}}}
%%%%%%%%%%%%%%%%%%%%%%%%%%%%%%%%%%

\def\nn{\\  \nonumber}
\def\de{\partial}
\def\brf{{\mathbf f}}
\def\bbf{\bar{\bf f}}
\def\bF{{\bf F}}
\def\bbF{\bar{\bf F}}
\def\bA{{\mathbf A}}
\def\bB{{\mathbf B}}
\def\bG{{\mathbf G}}
\def\bI{{\mathbf I}}
\def\bM{{\mathbf M}}
\def\bY{{\mathbf Y}}
\def\bX{{\mathbf X}}
\def\bS{{\mathbf S}}
\def\bb{{\mathbf b}}
\def\bh{{\mathbf h}}
\def\bg{{\mathbf g}}
\def\bla{{\mathbf \la}}
\def\bmu{\mathbf m }
\def\by{{\mathbf y}}
\def\bmu{\mbox{\boldmath $\mu$} }
\def\bsig{\mbox{\boldmath $\sigma$} }
\def\bunity{{\mathbf 1}}
\def\cA{{\cal A}}
\def\cB{{\cal B}}
\def\cC{{\cal C}}
\def\cD{{\cal D}}
\def\cF{{\cal F}}
\def\cG{{\cal G}}
\def\cH{{\cal H}}
\def\cI{{\cal I}}
\def\cL{{\cal L}}
\def\cN{{\cal N}}
\def\cM{{\cal M}}
\def\cO{{\cal O}}
\def\cR{{\cal R}}
\def\cS{{\cal S}}
\def\cT{{\cal T}}
\def\eV{{\rm eV}}

%
%%%%%%%%%%%%%%%%%%%%%%%%%%%%%%%%%%%%%

\title{Brane Bounce from logarithmic entropic corrections in the bulk}

\author{Andrea Addazi$^1$}\email{andrea.addazi@infn.lngs.it}
\affiliation{$^1$ Dipartimento di Fisica,
 Universit\`a di L'Aquila, 67010 Coppito AQ and
LNGS, Laboratori Nazionali del Gran Sasso, 67010 Assergi AQ, Italy}

\begin{abstract}

We calculate new corrections to the Brane-world dynamics, lying in a 5D Schwarzschild-De Sitter black hole, generalizing the 
result of Nojiri, Odintsov and Ogushi (NOO) in Ref.\cite{Nojiri:2002vu}, 
The NOO entropy effect is based on the Logharitmic correction
to the bulk entropy firstly 
calculated by  Mukherji and Pal in Ref.\cite{Mukherji:2002de}. 
We calculate higher order contributions 
to the brane worldsheet. 
The extra terms obtained lead to interesting implications in brane-cosmology. 
In particular, new entropic terms rapidly disappear in the late Universe
while exploding in the very Early Universe.
In particular, we show that they may trigger a cosmological bounce 
in the very early Universe. On the other hand, 
they contribute to the 
 cosmological expansion in the Late Universe. 
We also discuss a scenario in which the BLK anisotropies 
are washed-out, 
toward
a new Ekpyrotic Brane Cosmology.  

\end{abstract}

\maketitle
\section{Introduction}

The semiclassical analysis of Branes lying in a higher dimensional bulk
is crucially important for our understanding of string theory and holographic principles.
On the other hand, a scenario in which our Universe is a four dimensional brane 
in a higher dimensional bulk 
was largely explored in literature 
(see Refs.\cite{Maartens:2010ar,Brax:2003fv} for useful reviews on these subjects). 
In these models, extra-dimensions are not compactified in Calabi-Yau 
manifolds and the Standard Model particles are localized in the Brane.
Gravitational fields are {\it leaked} in all the entire higher dimensional space-time. 
Contrary to SM particles, obtained from open strings attached in the Brane-World, 
gravitons are lowest massless vibrations of closed strings propagating 
in the bulk. This can lead to new interesting modifications of General Relativity. 
Brane-World cosmology seems to be particularly interesting 
for a fundamental understanding of the cosmo-genesis.
Bounce cosmology, (Brane) inflation and phantom/quintessence dark energy 
may be naturally explained in these scenarios \cite{Maartens:2010ar,Brax:2003fv}. 
The simplest one is inspired by Randall-Sundrum scenario, in 
which the brane-world lies in a $AdS_{5}$-bulk \cite{Randall:1999vf}. 
RS model can be also studied using the holographic 
AdS/CFT conjecture \cite{Maldacena:1997re}.
RS scenario can be generalized 
to a Schwarzschild-De Sitter bulk
and dS/CFT principle. 
In particular, the brane can lie in a 5D Schwarzschild-De Sitter black hole
with an Hubble radius which is much larger than the brane
horizon. This model may be extended considering effects of
other large extra dimensions, among the $D=9+1$ predicted by string theory.
On the other hand, as shown by Hawking forty years ago, 
the evaporation of 
static black hole is an unavoidable consequence of euclidean semiclassical quantum gravity 
and quantum field theories in curved space-time \cite{Hawking:1974sw}. 
In light of dS/CFT correspondence, 
Cardy and Verlide have shown their entropy formula in Refs.\cite{Verlinde:2000wg,Cardy}. 
This result can be generalized considering thermal fluctuations in 
the euclidean path integral. 
As is well known, the euclidean path integral coincides
with the black hole partition function,
which can be considered in the  macro-canonical ensamble. 
Mukherji and Pal in Ref.\cite{Mukherji:2002de} have shown that 
in macro-canonical ensamble, the Cardy-Verlinde formula
is corrected to logarithmic corrections, 
controlled by the black hole specific heat. 
As a consequence, a backreaction effect of the bulk entropy fluctuations 
and the brane lying inside may be expected. 
Nojiri, Odintsov and Ogushi (NOO)  have shown that 
the entropy fluctuations of the bulk black hole
can provide a new correction to the Hubble rate of the FRW brane. 
In particular, they have shown that the entropic
effect is tiny in the late Brane-Universe, 
but it explodes in the early Brane-Universe \cite{Nojiri:2002vu} 
\footnote{See also  \cite{Nojiri:2001nr,Nojiri:2001nr,Nojiri:2001ae,Nojiri:2002yi,Odintsov:2002zi,Nojiri:2002qt,Nojiri:2001ut,Nojiri:2002hz}
for related contributions in these subjects.}.
In this paper, we will analyze and generalize the Nojiri, Odintsov and Ogushi (NOO) entropic effect.
In particular, we will calculate new contributions to the FRW brane equation. 
We will obtain new extra terms scaling as $a^{-4}Log[a^{3}], a^{-8}Log[a^{3}]$.
These are highly suppressed in the late Brane Universe. 
However, they will provide an important contribution nearby the Planck scale. 
This provides an interesting mechanism for a Bounce/Oscillating Cosmology
in the Brane Universe. 

%DA AGGIUNGERE

\section{Logarithmic corrections to the Brane-World dynamics  }

Let us consider the five dimensional Einstein-Hilbert action
\begin{equation}
\label{A}
S=\int d^{5}x \sqrt{-G}\left\{\frac{1}{16\pi G_{5}}R+\Lambda \right\}
\end{equation}
We consider its Schwarzschild-De Sitter solution
\begin{equation}
\label{B}
ds_{5}^{2}=-e^{2\rho}dt^{2}+e^{-2\rho}da^{2}+a^{2}g_{ij}dx^{i}dx^{j}
\end{equation}
$$e^{2\rho}=\frac{1}{a^{2}}\left(-\frac{a^{4}}{l^{2}}-\mu+\frac{k}{2}a^{2} \right)$$
$l$ curvature radius of SdS, $\Lambda=12/l^{2}$. 

The thermodynamical proprieties of the curved SdS space-time can 
be evaluated from the the partition function in the 
gran canonical ensamble:
\begin{equation}
\label{PartitionFunction}
Z(\beta)=\int e^{-\beta E}\rho(E) dE
\end{equation}
where $\rho(E)$ is the density of states, 
$\beta=1/T_H$, $T_H$ is the Hawking's temperature $(k_{B}=1)$.
One can use the standard thermodynamical relations
among the free-energy, the internal energy, the entropy and the partition function
as
\begin{equation}
\label{FTS}
F=E-T_HS,\,\,\,F=-T_H Log[Z]
\end{equation}
\begin{equation}
\label{ebetaF}
e^{-\beta F}=\int dE L e^{-\beta E+S(E)},\,\,\,\,\rho(E)=Le^{S(E)}
\end{equation}
where $L$ is a characteristic length scale. 
Using the saddle point approximation, the expansion around the equilibrium point $E_{0}$ is 
\begin{equation}
\label{Exapansion}
-\beta E+S(E)=-\beta E_{0}+S(E_{0})
\end{equation}
$$+\frac{1}{2}\beta^{2}B(E_{0})(E-E_{0})^{2}+O((E-E_{0})^{2})$$
$B(E_{0})$ is a dimensionless constant. 
So that, one can get 
\begin{equation}
\label{ebetaF2}
e^{-\beta F}=\sqrt{\frac{\pi L^{2}}{\beta^{2}B(E_{0})}}e^{-\beta E_{0}+S(E_{0})}=e^{-\beta E_{0}+S(E_{0})+\frac{1}{2}Log \frac{\pi L^{2}}{\beta^{2}B(E_{0})}}
\end{equation}
leading to 
\begin{equation}
\label{betaF3}
-\beta F=-\beta E_{0}+S(E_{0})+\frac{1}{2}Log\left[\frac{C(E_{0})L^{2}}{\beta^{2}} \right]
\end{equation}
where $C(E_{0})=\pi/B(E_{0})$.
The specific heat is given by 
\begin{equation}
\label{specificheat}
\left[\frac{\partial \langle E \rangle}{\partial T}\right]_{V}=\beta^{2}\left( \langle E^{2}\rangle-\langle E \rangle^{2}\right)=\frac{1}{B(E_{0})}
\end{equation}
As a consequence the 0th order entropy is corrected as
\begin{equation}
\label{correction}
S(E_{0})=S_{0}-\frac{1}{2}Log\left[\frac{C(E_{0})L^{2}}{\beta^{2}} \right]
\end{equation}

The action on the FRW brane is 
\begin{equation}
\label{SW3T}
S=\frac{W_{3}}{T_H}\frac{1}{2\pi G_{5}l^{2}}\int_{a_{H}}^{\infty}da \,a^{3}
\end{equation}
renormalized as 
$$S=\frac{W_{3}}{T_H}\frac{1}{2\pi G_{5}l^{2}}$$
$$\times \left[\int_{a_{H}}^{a_{max}}da\, a^{3}-\left(e^{\rho(a_{max})-\rho(a_{max},\mu=0)}\right)\int_{0}^{a_{max}}da \,a^{3}\right]$$
According to Cardy-Verlinde formula \cite{Verlinde:2000wg,Cardy}, Eq.(\ref{SW3T}) implies a leading entropy as follows 
\begin{equation}
\label{S0FT}
S_{0}=-\frac{dF}{dT_H}=\frac{W_{3}a_{CH,BH}^{3}}{4G_{5}}
\end{equation}
(both cosmological horizon and black hole),
$$T_{H}=\frac{1}{\pi}\left(\frac{W_{3}}{4G_{(5)}} \right)\left(1+\frac{2a_{H}}{l^{2}} \right)S_{0}^{-1/3}$$
$$F=\frac{W_{3}a^{2}_{BH}}{16\pi G_{5}}\left(1+\frac{a_{BH}^{2}}{l^{2}}\right),\,\,\, F=-\frac{W_{3}a^{2}_{CH}}{16\pi G_{5}}\left(1+\frac{a_{CH}^{2}}{l^{2}}\right)$$
$$E=F+T_H S=\pm \frac{3W_{3}\mu}{16\pi G_{5}}$$
with $+$ corresponding to the BH and $-$ to the CH.  
The Hawking's temperatures for the two horizons are
\begin{equation}
\label{C}
T_{H}(a_{CH})=|de^{2\rho}/da|_{a=a_{CH}}=-\frac{1}{2\pi a_{CH}}+\frac{a_{CH}}{\pi l^{2}}
\end{equation}
\begin{equation}
\label{C}
T_{H}(a_{BH})=|de^{2\rho}/da|_{a=a_{CH}}=\frac{1}{2\pi a_{BH}}-\frac{a_{BH}}{\pi l^{2}}
\end{equation}
where $a_{CH,BH}$ are the cosmological and black hole horizons respectively
which read as 
$$a_{CH,BH}=kl^{2}\pm \frac{1}{2}\sqrt{k^{2}l^{4}-4\mu l^{2}}$$

From Eq.(\ref{correction}),
the Logharitmic correction of the Cardy-Verlinde entropy is \cite{Mukherji:2002de}
\begin{equation}
\label{S}
S=S_{0}-\frac{1}{2}log \,C_{v}+...
\end{equation}
where $C_{v}$ is the specific heat
$$C_{v}=\frac{dE}{dT_H}=3\frac{2a_{H}^{2}-l^{2}}{2a_{H}^{2}+l^{2}}S_{0}$$

The 4d energy of the FRW brane in SdS bulk
\begin{equation}
\label{E4}
E_{4}=\frac{lE}{a}=\pm \frac{3 W_{3}l \mu}{16 \pi G_{5}a}
\end{equation}
$+$ for BH, $-$ for Cosmological. 
This implies that the temperature, entropy and heat capacity on the brane 
are related to the bulk ones as 
$$T=\left( \frac{l}{a}\right)T_{H},\,\,\,\,S^{B}_{0}=\left(\frac{a}{l} \right)^{3}S_{0},\,\,\,C_{V}^{B}=\left(\frac{a}{l} \right)^{3}C_{V}$$

The Casimir energy corresponds to 
\begin{equation}
\label{Casimir}
E_{C}=3(E_{4}+pW-TS)
\end{equation}
where $p=E_{4}/3W$ is the pressure where $W=a^{3}W_{3}$. 
Eq.(\ref{Casimir}) with uncorrected entropy is 
\begin{equation}
\label{ECaH}
E_{C}=\pm \left(\frac{3l a^{2}W_{3}}{8\pi G_{5}a} \right)
\end{equation}
The entropy shift
$$S_{0}\rightarrow S_{0}-\frac{1}{2}Log C_{v}$$
corresponds to an brane energy shift
\begin{equation}
\label{deltaET}
\delta E_{(4)}=-\frac{T}{2} Log[C_{v}]
\end{equation}
and so that to a shift in the energy density 
entering in the FRW brane: 
$$\frac{8\pi G_{4}}{3}\frac{\delta E_{(4)}}{W}=-\frac{8\pi G_{4}}{3}\frac{T}{2W_{3}a^{3}} Log[ C_{v}]$$

The induced metric $\gamma$ on the brane
$$\gamma_{ab}=G_{\mu\nu}e^{\mu}_{a}e^{\nu}_{b}$$
where $g_{\mu\nu}$ is the Bulk metric, 
$e_{a}^{\mu}$ are the vectors spanning the tangent space of 
the brane $\Sigma$. 
The 
Parallel transports on the brane 
and the bulk are 
$$u^{a}D_{a}u^{c}=0,\,\,\,\,u^{\mu}\nabla_{\mu}u^{\nu}=0,\,\,\,\,u^{\mu}=u^{a}e_{a}^{\mu}$$
The relations among parallel transports are 
\begin{equation}
\label{deltaET}
u^{\mu}\nabla_{\mu}u^{\nu}=(u^{a}D_{a}u^{c})e_{c}^{\nu}-K_{ab}u^{a}u^{b}n^{\nu}
\end{equation}
where $K_{ab}$ is the extrinsic curvature of $\Sigma$ and $n^{\mu}$
is the normal vector of $\Sigma$. 
Brane in the bulk is like a jump in its extrinsic curvatures
$K^{+}_{ab},K^{-}_{ab}$. They satisfy the junction condition: 
$$K_{ab}^{+}-K_{ab}^{-}=-\kappa_{5}^{2}\left(S_{ab}-\frac{1}{3}S \gamma_{ab}\right)$$
where $S_{ab}$ is the surface energy-momentum tensor of the brane. 
One can impose the $Z_{2}$ symmetry with respect to
the surface $\Sigma$:
$$K_{ab}^{+}=-K_{ab}^{-}$$
Form the junction condition with $Z_{2}$ symmetry 
with respect to $\Sigma$ for the surface energy-momentum tensor, 
we can obtained the extended FRW equation for the brane: 
\begin{equation}
\label{Hubble}
H^{2}=-\frac{1}{l^{2}}+\frac{k}{a^{2}}-\frac{8\pi G_{(4)}}{3}\rho
\end{equation}

$$-l^{2}\left( \frac{4\pi G_{(4)}}{3}\rho\right)^{2}
+\frac{4\pi G_{(4)} T(a)}{3 W}Log \,C_{v}^{B}(a)$$
$$-l^{2}\left( \frac{2\pi G_{(4)} T(a)}{3 W}Log \,C_{v}^{B}(a)\right)^{2}$$
$$+2\left(\frac{8\pi G_{(4)}}{3}\rho\right) \left(\frac{2\pi G_{(4)} lT(a)}{3 W}Log \,C_{v}^{B}(a)\right)$$
with
$$\rho=\rho_{(0)}+\rho_{m}$$
$$\rho_{(0)}=\frac{E_{4}}{W}=\frac{3l\mu}{16\pi G_{5} }\frac{1}{a^{4}}$$
$$G_{(4)}=\frac{2G_{(5)}}{l}, \,\,\,T(a)=\frac{l}{a}T_{H}$$
 $\rho_{m},\rho_{\Lambda}$ are matter density, localized in the FRW brane, and vacuum energy density,
 respectively. 
This equation generalizes the extended FRW
equations obtained in Refs. \cite{EFRW1,EFRW2,EFRW3,EFRW4}.

In the following analysis, we will consider the case of a brane expanding 
in bulk and never entering in the Schwarzschild horizon. 

\vspace{0.3cm}
Eq.(\ref{Hubble}) corresponds to
\begin{equation}
\label{Conservation}
\dot{a}^{2}+V(a)=k
\end{equation}
where 
\begin{equation}
\label{Hubble}
V(a)=c_{0}a^{2}-c_{1}a^{-2}Log[\gamma_{1}a^{3}]+
c_{2}a^{-2}
\end{equation}
$$+c_{3}a^{-6}\left(Log[\gamma_{1}a^{3}]\right)^{2}
+c_{4}a^{-6}-c_{5}a^{-6}Log[\gamma_{1}a^{3}]$$
%$$+c_{6}a^{2}Log[\gamma_{1}a]$$ ???
assuming $\rho_{m}=0$. 
$$c_{0}=\frac{1}{l^{2}},\,\,\,c_{1}=\frac{4\pi G_{(4)} lT_{H}}{3 W_{3}}$$
$$c_{2}=\left(\frac{8\pi G_{(4)}}{3}\right)\left(\frac{3l\mu}{16\pi G_{5} }\right)$$
$$c_{3}=l^{2}\left( \frac{\pi G_{(4)} T_{H}l}{3 W_{3}}\right)^{2}$$
$$c_{4}=l^{2}\left(\frac{ G_{(4)}}{3}\frac{l\mu}{2\pi G_{(5)} }\right)^{2}$$
$$c_{5}=c_{1}c_{2},\,\,\,\gamma_{1}=\frac{C_{V}}{l^{3}}$$

Eq.(\ref{Conservation}) is equivalent to the one-dimensional 
motion of a particle. 
In Fig.1, we show the opportunely normalized brane potentials 
scanning on the space of parameters. 
In particular, for a brane starting from $a=\infty$, with $k=0,1$, and approaching 
$a\rightarrow 0$, it can reach a maximal value $a_{max}$
and return back, re-expanding. 
On the other hand, for a brane starting from $a=0$, 
it can reach the maximum $a_{max}$ and it can re-collapse back to $a=0$. 

 \begin{figure}[t]
\centerline{ \includegraphics [height=5cm,width=0.7 \columnwidth]{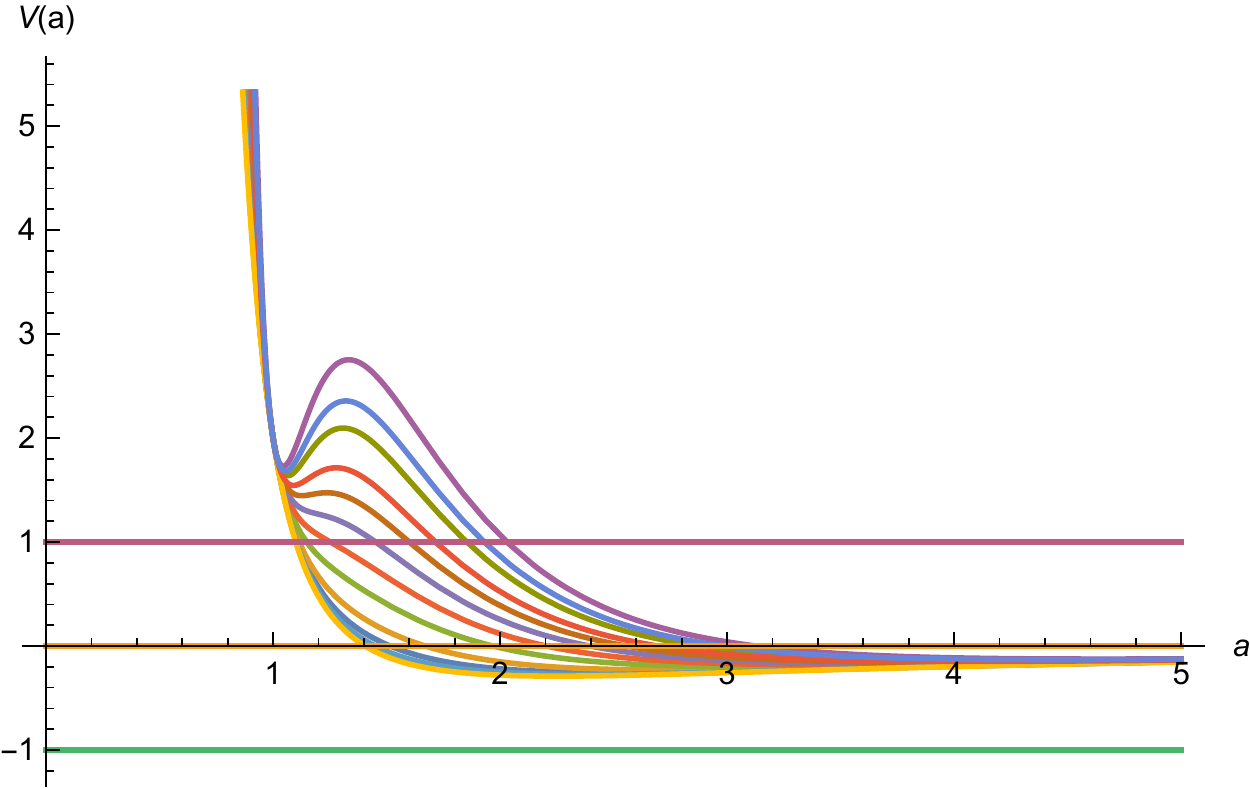}}
\vspace*{-1ex}
\caption{The Brane potentials $V\equiv V(a)$ are displayed
with several choices of the parameters
(the unit convention $c_{1,2,4,5}\gamma_{1}=1$
varying $c_{3}$). Brane turning-points levels 
$k=-1,0,1$ are also displayed. 
}
\label{plot}   % \ref{plot}
\end{figure}

\section{Toward a Realistic Ekpyrotic Brane Cosmology}

In the Bounce scenario triggered by the NOO effect, 
the growing of anisotropies in the energy-momentum tensor
are expected to be order $O(1)$. This is a typical problem 
in every Bounce mechanisms: the 
new Universe would emerge from the BLK (Belinsky, E. M. Lifshits and I. M. Khalatnikov)
singularity from the pre-Universe \cite{BLK1,BLK2,BLK3}. 
Of course this problem can be solved 
assuming 
a successive inflation mechanism. 
For example a brane inflation scenarios could over-implemented in 
our picture.
However, 
in this section, we will consider another solution
involving the introduction of an extra
exotic super-stiff fluids with $p_{\phi}>\rho_{\phi}$ \cite{Erickson:2003zm,Rubakov:2014jja}.
As is known, a way-out from the 
BLK anisotropies is that the Early Universe 
is dominated by the overall condition $p_{\phi}>\rho_{\phi}$ \cite{Erickson:2003zm}.
However, in our case, the introduction of the exotic superstiff fluid 
introduce a new interesting term $\rho_{\phi}a^{-4}Log[a^{-3}]$. 
Such a term is repulsive in the late Universe,
while attractive in the very Early Universe. 
An example of superstiff fluid is a scalar field 
with an exponential potential:
\begin{equation}
\label{Lphi}
L_{\phi}=\frac{1}{2}\partial_{\mu}\phi \partial^{\mu} \phi-V(\phi),\,\,\,V(\phi)=-V_{0}e^{\phi/M}
\end{equation}
The corresponding energy density is positive and 
highly increasing with $a$ decreasing:
\begin{equation}
\label{rhoooo}
\rho_{\phi}=\frac{1}{2}\dot{\phi}^{2}+V(\phi)=\frac{6M^{2}\alpha}{t^{2}}=\frac{6M^{2}\alpha}{a^{2/\alpha}}
\end{equation}
while the pressure is 
\begin{equation}
\label{poooo}
p_{\phi}=\frac{1}{2}\dot{\phi}^{2}-V(\phi)=\frac{4M^{2}}{t^{2}}=\frac{4M^{2}\alpha}{a^{2/\alpha}}
\end{equation}
where $a(t)=|t|^{\alpha}$ and 
$w>1$ corresponds to $\alpha<1/3$. 
The presence of such a fluid generates an repulsive entropic term contributing to as $H^{2}$ as
\begin{equation}
\label{rhoooo}
2\left(\frac{8\pi G_{4}}{3}\rho_{\phi}\right) \left(\frac{2G_{4}}{3Wl}Log \,3S_{0}\right)\sim a^{2/\alpha} a^{-4}Log[(...)a^{3}] 
\end{equation}
which is minor than $<a^{-9}Log[(...)a]$. 
In Fig.2-3, we show the effect of such an entropic term with the growing of $M/M_{Pl}$
in the case $\alpha=1/4$ and $\mu=1$.
As we can see, the new term generated a potential barrier which
radically changes the bounce dynamics. 
As an alternative, we consider a repulsive superstiff exotic fluid, 
with the same potential of Eq.(\ref{Lphi}).
In Fig.4, the brane dynamics 
in presence of a repulsive superstiff fluid is displayed. 

 \begin{figure}[t]
\centerline{ \includegraphics [height=5cm,width=0.7 \columnwidth]{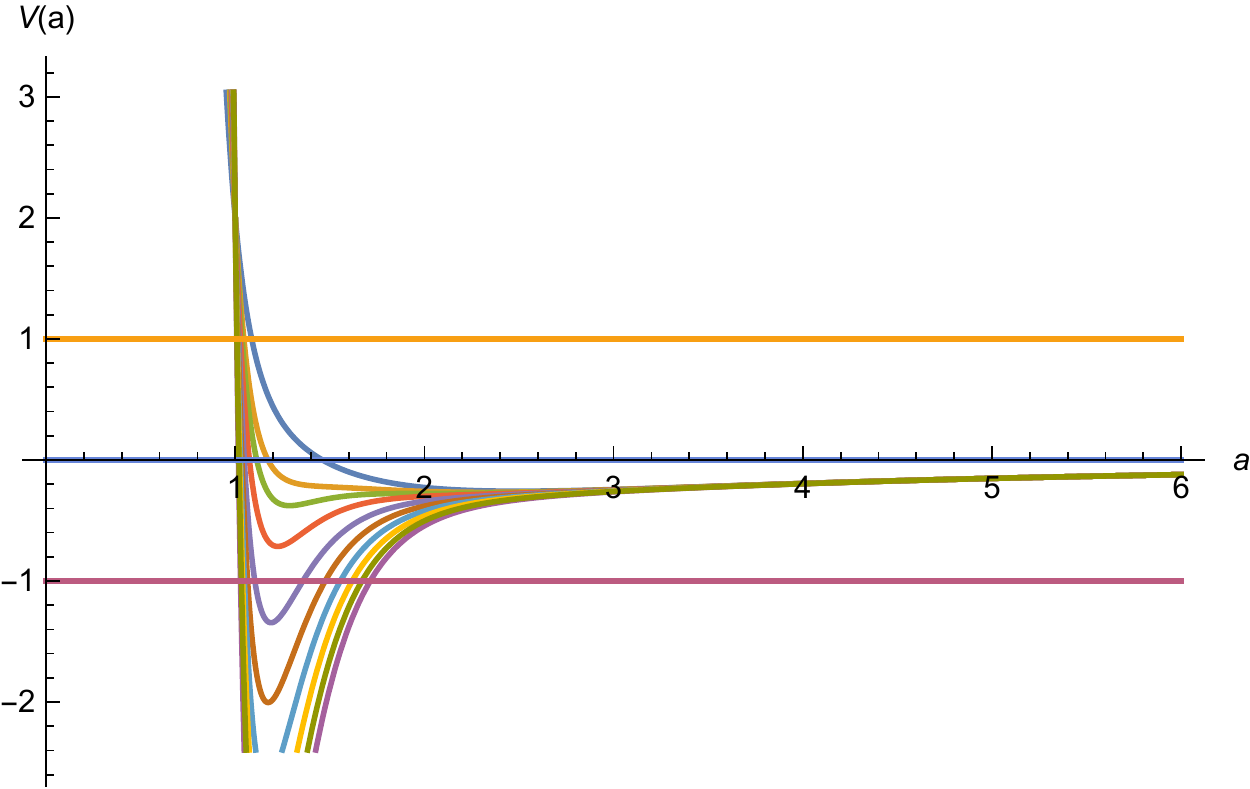}}
\vspace*{-1ex}
\caption{The Brane potentials $V\equiv V(a)$ 
in presence of superstiff fluids
are displayed. We chose $\alpha=1/4$,
with several choices of the parameters. 
(the unit convention $c_{1,2,3,4,5},\gamma_{1}=1$
varying the overall coefficient in Eq.(\ref{rhoooo}).
Brane turning-points levels 
$k=-1,0,1$ are also displayed. 
}
\label{plot}   % \ref{plot}
\end{figure}
 \begin{figure}[t]
\centerline{ \includegraphics [height=5cm,width=0.8 \columnwidth]{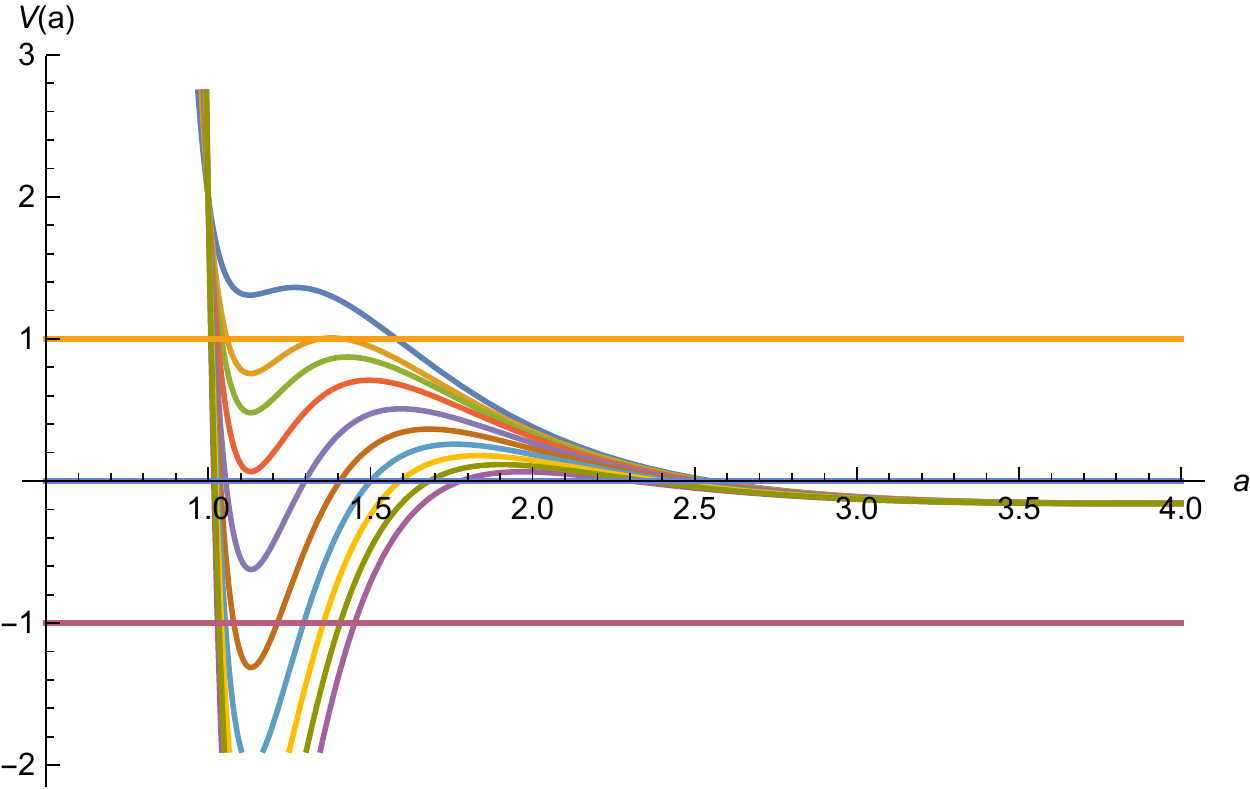}}
\vspace*{-1ex}
\caption{The Brane potentials $V\equiv V(a)$ 
in presence of superstiff fluids
are displayed. Brane turning-points levels 
$k=-1,0,1$ are also displayed.  }
\label{plot}   % \ref{plot}
\end{figure}

 \begin{figure}[t]
\centerline{ \includegraphics [height=5cm,width=0.8 \columnwidth]{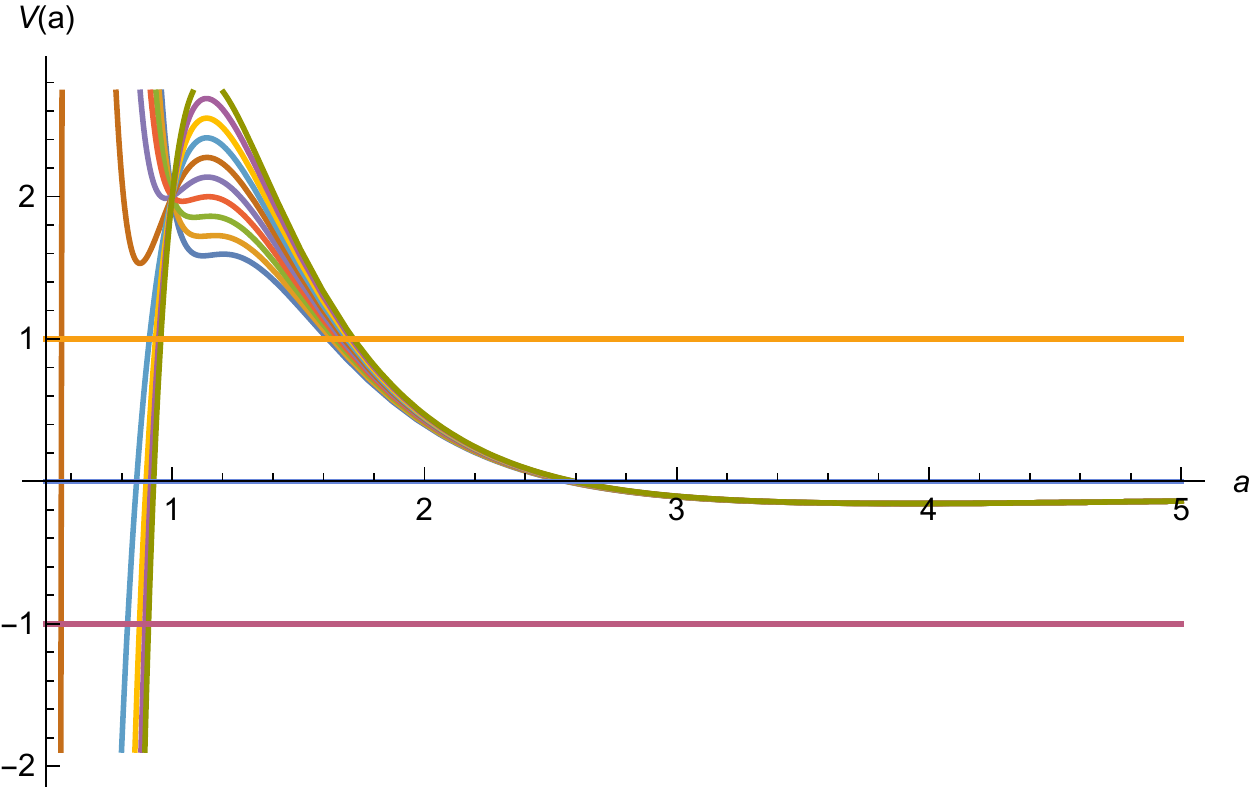}}
\vspace*{-1ex}
\caption{The Brane potentials $V\equiv V(a)$ 
in presence of superstiff fluids
are displayed. Brane turning-points levels 
$k=-1,0,1$ are also displayed.  }
\label{plot}   % \ref{plot}
\end{figure}

\section{Conclusions}
In this paper, we have shown how the thermal entropy fluctuations 
of a higher dimensional Schwarzschild-De Sitter black hole 
may provoke modifications of the Brane FRW dynamics. 
We have generalized the Nojiri-Odintsov-Ogushi  result, 
calculating a more complete equation for the brane 
lying in the Schwarzschild-De Sitter black hole. 
In particular, neq entropic terms 
are one scaling as $a^{-4}Log[a^{3}]$ and $a^{-8}(Log[a^{3}])$.
The first one can be interpreted as a {\it repulsive dark radiation}
in the Late Universe, while becoming attractive 
in the Early Universe $a\rightarrow 0$. 
The second class of terms provides highly suppressed 
contributions in the Late Universe, while exploding for $a\rightarrow 0$. 

This implies that entropic terms may relevantly change the dynamical behavior of the brane 
in the early Universe and in the very late Universe. 

On the other hand, we also find terms scaling as $\rho_{m}a^{-4}Log[a^{3}]$ 
where $\rho_{m}$ is the matter density localized 
in the four dimensional brane. As a consequence, the fate of the brane 
Universe is highly dependent by the matter and vacuum energy content in it and in the bulk. 
These new matter/vacuum dependent terms are fundamentally important 
in realistic Ekpyrotic Brane Cosmology.
As a useful example, we have shown the case of a supestiff 
fluid which washes out BLK anisotropies in the bounce. 
In this case, the superstiff fluid also triggers the Bounce mechanism 
thanks to the associated extra entropic repulsive term. 
As a consequence, the superstiff fluid does not only 
solve the BLK anisotropy problem but it 
is also a source of the Bounce super-repulsion. 

Let us also comment on 
the possible presence 
of cosmological terms induced by 
dark Yang-Mills condensates 
or dark non-linear Born-Infeld electrodynamics. 
In the framework of dark Yang-Mills or dark QCD condensates,
cold dark matter and dark energy can be elegantly 
unify in a common framework, as recently shown in Refs.\cite{Addazi:2016sot,Addazi:2016nok}. 
On the other hand, if the Dark Born-Infeld sector 
is coupled to neutrinos, a order $10^{-3}\, \rm eV$ neutrino mass 
can be generated and connected with dark energy \cite{Addazi:2016oob}. 
In both these scenarios the repulsive condensates 
may change in cosmological time and 
their associated entropic terms may 
drive 
the early cosmological evolution 
in a complicated way. 
A complete analysis of these cases 
deserves further investigations beyond the 
purposes of this letter.
Finally, let us note that the entropic corrections may imply interesting unexplored effects
in more complicated Brane-world scenarios. 
For example in 
$f(R)$-Brane worlds, 
interesting amplifications or suppressions of the entropic terms
would be obtained.
On the other,
in Extended Theories of gravity, the thermodynamical behavior of 
black hole may be highly non-trivial.
For instance,  as discussed in Ref.\cite{Addazi:2016prb},
the Hawking's emission is exponentially suppressed 
 for Bousso-Hawking-Nojiri-Odintsov
antievaporating solution   \cite{Bousso:1997wi,Nojiri:2013su,Nojiri:2014jqa}
\footnote{This effect was recently studied in context of string-inspired black holes
\cite{Addazi:2016hip}. }.

\vspace{0.5cm} 

{\large \bf Acknowledgments} 
\vspace{3mm}

My work was supported in part by the MIUR research grant Theoretical Astroparticle Physics PRIN 2012CP-PYP7 and by SdC Progetto speciale Multiasse La Societa della Conoscenza in Abruzzo PO FSE Abruzzo 2007-2013.

\end{document}